\documentclass[journal = jctcce, manuscript = article, layout = twocolumn ]{achemso}

\setkeys{acs}{email = false}
\usepackage[utf8]{inputenc}
\usepackage{graphicx}
\usepackage{amsmath}
\usepackage{bm}
\usepackage{microtype}
\SectionNumbersOn

\usepackage{enumitem,booktabs}
\usepackage[referable]{threeparttablex}
\renewlist{tablenotes}{enumerate}{1}
\makeatletter
\setlist[tablenotes]{label=\tnote{\alph*},ref=\alph*,itemsep=\z@,topsep=\z@skip,partopsep=\z@skip,parsep=\z@,itemindent=\z@,labelindent=\tabcolsep,labelsep=.2em,leftmargin=*,align=left,before={\footnotesize}}
\makeatother

\title{Constant Size Molecular Descriptors For Use With Machine Learning}

\author{Christopher R. Collins}
\affiliation[CMUChem]{Department of Chemistry, Carnegie Mellon University, Pittsburgh, Pennsylvania 15213 United States}
\author{Geoffrey J. Gordon}
\affiliation[CMUML]{Machine Learning Department, Carnegie Mellon University, Pittsburgh, Pennsylvania 15213 United States}
\author{O. Anatole von Lilienfeld}
\affiliation[University of Basel]{Department of Chemistry, Institute of Physical Chemistry and National Center for Computational Design and Discovery of Novel Materials (MARVEL), University of Basel, 4056 Basel, Switzerland}
\author{David J. Yaron}
\email{yaron@cmu.edu}
\affiliation[CMUChem]{Department of Chemistry, Carnegie Mellon University, Pittsburgh, Pennsylvania 15213 United States}

\newcommand{\bestf}{\mbox{$\bm{1 2^{NP} 3^{B} 4^{B}}$}}
\newcommand{\simplef}{\mbox{$\bm{1 2^{NP} 3^{B}}$}}
\newcommand{\fv}[1]{\mbox{$\bm{#1}$}}
\DeclareSymbolFont{bbold}{U}{bbold}{m}{n}
\DeclareSymbolFontAlphabet{\mathbbm}{bbold}
\begin{document}

\begin{abstract}
A set of molecular descriptors whose length is independent of molecular size is developed for machine learning models that target thermodynamic and electronic properties of molecules. These features are evaluated by monitoring performance of kernel ridge regression models on well-studied data sets of small organic molecules. The features include connectivity counts, which require only the bonding pattern of the molecule, and encoded distances, which summarize distances between both bonded and non-bonded atoms and so require the full molecular geometry. In addition to having constant size, these features summarize information regarding the local environment of atoms and bonds, such that models can take advantage of similarities resulting from the presence of similar chemical fragments across molecules. Combining these two types of features leads to models whose performance is comparable to or better than the current state of the art. The features introduced here have the advantage of leading to models that may be trained on smaller molecules and then used successfully on larger molecules.
\end{abstract}

\maketitle

\section{Introduction}
    
Cheminformatics has a long tradition of using the techniques of machine learning to predict properties of molecules. A particular focus has been properties that emerge from interactions of the molecule with complex external systems, such as arise in drug discovery\cite{qsar,baldi_solubility}. More recently, informatics has been successfully used to develop models of the properties of the molecules themselves. This includes thermodynamic properties\cite{early_delta,coulomb_mat,atomization_review,coulomb_nips,soap,split_energies,ex_chem_screen2,nn_review,local_coulomb,scattering}, electronic\cite{clean_energy,clean_energy2,genetic,coulomb_multi_prop,delta_machine,delta_ml_electronic,independent_kernels,fr_coulomb,coulomb_forces,complex_props,resoap,systematic,ex_chem_screen,material_cartography} properties and molecular structure\cite{baldi_structure,baldi_sidechain,BASC,kernel_derivative}. Successful predictions of such molecular properties provides an alternative to quantum chemical computations, leading to useful predictions at a small fraction of the cost of quantum computations\cite{coulomb_mat,atomization_review}. Because the target properties involve only the molecule itself, as opposed to interactions with complex external systems, the descriptors that bring the most explanatory power to the informatics tasks may differ from those developed for targets involving external systems. This paper explores a number of descriptors, or features, for predicting the properties of individual molecules.

A key dimension along which informatics models of molecular properties differ is the degree to which results from chemical computations are used as features for the prediction task. On one end of this spectrum, the model corrects predictions from standard computational chemistry methods, so that they better match more expensive calculations\cite{ediz_yaron,delta_machine,delta_ml_electronic} or experiments\cite{early_delta,nn_exp}. At the other end of the spectrum, the only information used is what one might get from a line drawing or SMILES\cite{smiles} string. At the latter level, there have been many examples of using simple counts of atoms and bonds to make predictions of molecular properties\cite{bag_of_bonds,atomization_review}. The low computational cost of such models makes them useful for searching molecular databases\cite{database}. Their restricted applicability to potential energy minima or known bonding patterns; however, reduces their predictive power.

Between the above two extremes are models that utilize the 3D molecular geometry to make predictions. Models that use features based on the Coulomb matrix, such as the recent Bag of Bonds (BoB) feature vector, have been shown to be highly effective with this level of information\cite{bag_of_bonds,atomization_review}. However, the length of these feature vectors scales formally as the square of the number of atoms in the largest molecule of interest\cite{atom_comment}. There have been attempts to rectify this issue, but they have been unable to achieve the same level of performance\cite{fr_coulomb}. 

A primary goal of the work presented here is to develop feature vectors whose length is independent of molecular size and that subsequently lead to models that generalize well from smaller to larger systems. This should be possible if we assume: a) an effective chemical periodicity where large systems consist of smaller fragments which resemble exemplary chemistries present in the data base used for training, and b) locality implying additivity. For certain systems, these assumptions can break down; in particular, the latter one when strong electron delocalization effects are present as in metals or in the $\pi$-bonds of conjugated polymers. To achieve our goal, two classes of features are explored. 

The first class of features are based on connectivity counts, which have been explored in a number of different ways.\cite{split_energies,new,atom_walks,circular,graph_kernel_chem} Here, these features include counts of atom types and bond types, which we refer to as rank-1 and rank-2 features since they summarize patterns related to one and two atoms, respectively. Higher-rank features, that count occurrences of patterns related to three or more bonded atoms, are also explored. 

The second class of features are encoded distances. These summarize the pairwise distances between atoms in a molecule through a feature vector whose length is independent of molecular size. The feature vector is a smoothed histogram of distances between atoms, including non-bonded atoms.

Both classes of features summarize information regarding the local environment of atoms and bonds in the molecule. The resulting feature vectors therefore contain information on the fragments present in the molecule. In particular, when two molecules possess the same molecular fragment, the interactions amongst the atoms in that fragment make similar contributions to the feature vector. Use of these feature vectors to predict similarity between two molecules, as in the kernel methods below, therefore incorporates similarity resulting from the presence of similar fragments. In addition, the feature vectors include information regarding interactions amongst the fragments. The features may therefore enable models to generalize from smaller to larger molecules. 

\section{Related Work}

The features studied here are somewhat related to features developed for neural net models of potentials\cite{parrinello,nn_review,gastegger2015,gastegger2016}. In those models, the total energy is written as a sum over atomic energies, with the  atomic energies being functions of the local environment\cite{distance_cutoff}. The environment is described using features based on two and three body correlation functions around the atom. The encoded distance features explored here may be viewed as smoothed versions of two-body correlation functions between each pair of elements. We note that the latter lack uniqueness, a crucial feature property for guaranteeing error convergence with training set size\cite{fr_coulomb}. In particular, they will fail to distinguish homometric molecules.

Below, we use kernel-based learning methods to explore the utility of the features introduced here. Our studies therefore have some similarities to the Regularized Entropy Match Smooth Overlap of Atomic Positions (REMatch-SOAP) kernel, which provides an alternative means for predicting molecular properties using only 3D molecular geometries\cite{resoap}. REMatch-SOAP does not introduce a molecular feature vector and, instead, provides a means to directly compute the similarity between two molecular structures, as needed for kernel methods. The approach begins by computing the similarity between each of the atoms in the two molecules being compared, where the similarity measure includes the molecular environment. The molecular similarity is then obtained from a regularized entropy match of the resulting atomic similarities.

In contrast to REMatch-SOAP, the features introduced here are not limited to kernel methods and, when used in kernel methods, allow us to separately consider the influences of three factors: the feature vector, the distance metric used to connect features to the similarity between molecules, and the kernel function. Our emphasis here is on the features themselves. We use cross validation to select the kernel, including the width parameter for an exponential kernel function and the choice of a Laplacian kernel, which uses an L1 distance metric, versus a Gaussian kernel, which uses an L2 distance metric. The performance of the resulting models on well studied data sets is examined, with particular attention paid to the ability of the method to scale between different size molecules. 

\section{Data Sets}\label{sec:data}

This work compares performance of various molecular features on data sets that have been the subject of past work, demonstrating the predictive power of machine learning models to predict molecular properties. These data sets utilize molecules selected from the following two data sets, each of which contains only the bonding pattern as summarized in a SMILES string. The {\bf GDB13} data set contains 977 million molecules made of C, H, O, N, S, and Cl, with up to 13 heavy atoms\cite{gdb13}, and the {\bf GDB17} data set contains 166 billion molecules made up of C, H, O, N, and F with up to 17 heavy atoms\cite{gdb17}.

The {\bf QM7} data set corresponds to a subset of the GDB13 molecules, consisting of all 7,165 molecules that contain 7 or fewer heavy atoms of elements C, N, O, or S for which optimized structures and atomization energies were calculated with the Perdew-Burke-Ernzerhof hybrid functional (PBE0)\cite{coulomb_mat}. Models for the atomization energies in QM7 have been extensively studied, primarily using features derived from the Coulomb matrix\cite{coulomb_mat,atomization_review,coulomb_nips,bag_of_bonds,local_coulomb}. As such, this data set provides a useful benchmark for testing models and comparing to previous work.

The {\bf QM7b} data set extends the QM7 data set to molecules containing chlorine, and includes a broader range of properties such as HOMO and LUMO energies, excitation energies, and polarizabilities calculated at various levels of theory\cite{coulomb_multi_prop}. QM7b allows performance of machine learning models to be explored on a wider variety of properties. Past works have also used QM7b to explore multi-target regression methods which, by simultaneously predicting various properties, can potentially benefit from relationships between the target properties\cite{coulomb_nips,bag_of_bonds,coulomb_multi_prop}.

The {\bf QM9} data set\cite{qm9} contains data for a subset of the GDB17 molecules, consisting of 133,885 molecules that contain nine or fewer heavy atoms. (The name QM9 is used here to be consistent with that of QM7.) The QM9 data set has been used in several studies\cite{bag_of_bonds,independent_kernels,coulomb_forces,fr_coulomb} and includes optimized structures and 18 different properties calculated using B3LYP/6-31g(2df,p). Below, QM9 is used to explore the degree to which models training on smaller molecules can transfer to larger molecules. The large amount of data in QM9 also allows more extensive studies on the degree to which inclusion of additional data improves model performance.

\section{Machine learning}\label{sec:ml}

A variety of machine learning methods have already been used to predict molecular properties, including linear regression, linear ridge regression, support vector regression, kernel ridge regression, regression trees, k-nearest neighbors, and neural nets/deep learning\cite{coulomb_multi_prop,atomization_review,nn_review}. For this study, we have elected to use two standard machine learning methods\cite{scikit-learn}, linear ridge regression (LRR) and kernel ridge regression (KRR) with the Gaussian and Laplacian kernels. These methods were chosen due to their relative simplicity, compared for instance to deep learning, and because past work has shown them to be effective on the data sets studied here\cite{atomization_review,bag_of_bonds}.

The models involve very many parameters which, during ``model training'', are adjusted to obtain predictions that agree with a given set of examples. In addition, both LRR and KRR involve hyperparameters that serve to define further the model form. The wide range of values for the hyperparameters leads to a wide range of possible models. ``Model selection'' involves searching through these to find a model, or equivalently a set of hyperparameters, that leads to good performance. Here, we do an exhaustive search over a grid of hyperparameter values using cross validation as described below. 

In LRR, there is one hyperparameter, $\alpha$, which is the weight on the ridge term.  When $\alpha$ is zero, the model produces the same result as linear regression. Finite values for $\alpha$ penalize larger regression parameters and so using finite values for $\alpha$ favors simpler models (those with smaller regression parameters). For LRR, the following $\alpha$ values were scanned over: $\{10^{-5}, 10^{-3}, \ldots, 10^{1}\}$. KRR retains the $\alpha$ parameter of LRR and adds two more.  The choice of kernel for KRR can be considered as a discrete hyperparameter and here the Gaussian ($\exp(-\gamma ||x-y||_2^2)$) and Laplacian ($\exp(-\gamma ||x-y||_1)$) kernels are included in the search.  For each of these kernels, there is an additional continuous hyperparameter, $\gamma$, which sets the width of the kernel. The search over $\alpha$ and $\gamma$ is exhaustive on a square grid with values $\{10^{-11}, 10^{-9}, \ldots, 10^{-1}\}$.

Model selection and testing use $k$-fold cross validation (CV). In $k$-fold CV, the examples in the data set are split into $k$ equally sized sets. One of these folds is then held out, and the examples in the remaining folds are used to train a model. The mean absolute error (MAE) for predictions made on the examples in the hold out set is then computed. This process is repeated, with each one of the $k$ folds being the hold out set, leading to $k$ different MAEs. The average of these MAEs is then taken as a measure of model performance. 

Here, model selection is based on a loop over $k$=5 splits. Within this loop, one split is ignored and the remaining 80\% of data is used to evaluate each set of hyperparameters, via 5-fold cross validation. This generates 5 figures of merit for each set of hyperparameters. The hyperparameters that have the best average performance are then selected as the final model. The MAE for this model, reported below and in the Supporting Information, is the average from a final 5-fold cross validation performed on the entire data set, using the selected hyperparameters. (See Supporting Information for the hyperparameters of each model.)

\section{Features}\label{sec:features}

The primary goal of this work is a comparison of models developed using different types of molecular features. Ideally, identical molecular geometries should lead to identical feature vectors, implying that the features are invariant with respect to molecular translations and rotations, and to reordering of the atoms. 

For models trained on smaller molecules to transfer to larger molecules, a number of other properties of the feature vectors are desirable. One such property is that the length of the feature vector be independent of molecular size. For the Coulomb matrix and BoB features, this property does not hold, as the number of elements in the feature vector scales quadratically with the number of atoms in the largest molecule in the data set. To support models of differently sized molecules, such feature vectors are expanded to a length that can accommodate the largest molecules in the data set, with zeros added to pad the extended regions for smaller molecules. For the remaining features considered below, the feature vectors have a length that depends on the molecular diversity, e.g. atom and bond types, but does not depend on molecular size. 

\subsection{Null model}\label{sec:null}

A ``null'' model is used to provide a baseline measure of the difficulty of the prediction task. The null model always predicts the mean value of the training data. The MAE of the null model therefore reflects the spread of the data.

\subsection{Coulomb Matrix and Bag of Bonds}\label{sec:cm}

A class of features that has led to well-performing models of molecular properties are those derived from the Coulomb matrix\cite{coulomb_mat,atomization_review,coulomb_nips,coulomb_multi_prop,delta_machine,delta_ml_electronic,independent_kernels,coulomb_forces},
\begin{equation}
M_{ij} =
\begin{cases}
    0.5 Z_i^{2.4}, & \text{if } i=j \\
    \frac{Z_i Z_j}{|r_i - r_j|}, & \text{if } i \neq j
\end{cases}
\label{eq:coulomb_matrix}
\end{equation}
where $Z_i$ and $r_i$ are the atomic number and Cartesian position of the $i^{th}$ atom. This feature is appealing in that it summarizes the molecular structure in terms of Coulomb interactions between atoms. Alternative forms for the interaction, other than Coulomb interactions between bare nuclei, have been tried but were not found to enhance model performance\cite{bag_of_bonds}. 

While the Coulomb matrix is invariant to molecular rotations and translations, it is not invariant to reordering of the atoms. A number of means have been explored to address the dependence on atom order, including sorting the matrix based on the magnitude of the norm of the rows, using randomly permuted sets of matrices, and using the eigenvalues of the matrix\cite{atomization_review,coulomb_nips,coulomb_multi_prop}. (The results generated here do not sort the Coulomb matrix, and simply arrange the lower triangle of the symmetric matrix into a vector.) A recent successful approach rearranges the Coulomb matrix into a ``Bag of Bonds'', BoB, feature \cite{bag_of_bonds}. In BoB, off-diagonal elements of the matrix are first divided into bags, based on the elements involved in the Coulomb interaction (CH, CC, CO, etc). The values in each bag are then sorted from high to low values. The maximum size of each bag, across the molecules in the data set, is then determined and zeros are added to each bag so that all molecules have bags of the same length. The bags are then concatenated to make a single feature vector that is invariant to reordering of the atoms. Because the BoB model is just a reordering of Coulomb matrix elements, the length of the vector still grows quadratically with the size of the largest molecule in the data set. BoB has also recently been extended to include three-body and higher terms\cite{BAML}, leading to feature vectors who length grows as the third or higher power of the number of atoms in the largest molecule.

\subsection{Connectivity Counts}\label{sec:conn}

This class of features summarizes the bonding pattern of a molecule through features that count occurrences of some pattern regarding the bonding between atoms. Each feature may be assigned a rank, based on the number of distinct atoms that are examined when testing for the presence of the pattern.  Rank-1 features count atom types, such as element or element plus coordination number, leading to a vector whose length is equal to the number of atom types. Rank-2 features count bond types (single, aromatic, double, triple), leading to a vector whose length is the number of bond types. Rank-3 features count triplets of bonded atoms, and so on.

For rank 1, each element of the feature vector, $v_T$, holds the number of atoms in the molecule that have type $T$,
\begin{equation}
  v_{T} = \sum_i \delta_{A(i), T}
\end{equation}
where the sum is over atoms, $i$, $A(i)$ is a function that returns the atom type, and $\delta$ is the Kronecker symbol. Two atom typing schemes are explored here, leading to two different rank-1 feature vectors denoted as \fv{1} and \fv{1^C}.  For \fv{1}, atom type is defined by element only. For \fv{1^{C}}, atom type is defined by a combination of element and coordination number. Coordination number is taken as the number of bonds in which the atom participates.

For rank 2, each element of the feature vector, $v_T$, holds the number of bonds in the molecule that have type $T$, 
\begin{equation}\label{eq:chain2}
  v_T = \sum_{i<j} \delta_{B(i,j), T}
\end{equation}
where the sum is over atoms, and $B(i,j)$ is a function that returns the bond type between atoms $i$ and $j$, or null if a bond is not present. Two bond typing schemes are explored here, leading to two different rank-2 feature vectors denoted as \fv{2} and \fv{2^{B}}. For \fv{2}, bond type is defined by the two elements participating in the bond. A bond is assumed to be present if the separation between atoms is less than the cutoffs listed in the Supporting Information. For \fv{2^{B}}, bond type is defined by the two elements participating in the bond and the bond order, with bond orders assigned based on bond length as discussed in the Supporting Information.

For rank 3, each element of the feature vector corresponds to a tuple of two bond types, $(T_1,T_2)$, with a value
\begin{equation}\label{eq:rank3}
  v_{(T_1,T_2)} = \sum_{i<j<k} \delta_{B(i,j), T_1} \delta_{B(j,k), T_2}
\end{equation}
where the sum is over atoms. The feature counts only situations where atoms $i$ and $k$ are bonded to a common atom, $j$. The summation is thus over all triples of atoms to which a bond angle would typically be ascribed. Higher-rank patterns are generalizations of Eq.~\eqref{eq:rank3} to higher order. For example, the rank 4 summation is 
\begin{equation}\label{eq:rank4}
 v_{(T_1,T_2,T_3)} = \sum_{i<j<k<l} \delta_{B(i,j), T_1} \delta_{B(j,k), T_2}\delta_{B(k,l), T_3}
\end{equation}
with the summation being over all sets of four atoms to which a dihedral angle would typically be ascribed.

\subsection{Encoded distances}\label{sec:encoded}

The connectivity count features of Section~\ref{sec:conn} summarize the bonding pattern of a molecule but do not include information about the bond lengths, bond angles, or dihedral angles. The encoded features developed here allow information on the 3D structure of the molecule to be used while keeping the length of the feature vector independent of the size of the molecules included in the data set. We include only rank-2 encodings, which summarize information regarding atom-atom distances. Extension to higher ranks would lead to a substantial increase in the length of the total feature vector and are not explored here.

The encoded distance features may be viewed as generalizations of the above rank-2 connectivity features. Because the bond types of Eq.~\eqref{eq:chain2} are based on atom-atom distances, the rank-2 connectivity counts of Section~\ref{sec:conn} may be viewed as encoding atom-atom distances in a coarse grained manner. To introduce the notation we will use for encoding, we rewrite Eq.~\eqref{eq:chain2} as
\begin{equation}\label{eq:bond_count}
v_{T} = \sum_{j<k} \mathbbm{1}(r_{jk} \in (r_T^{low},r_T^{high}] ) 
\end{equation}
where $r_T^{low}$ and $r_T^{high}$ are the lower and upper limits of the distance ranges that define the bond type, $T$, for the corresponding pair of elements, and $r_{jk}$ is the distance between atoms $j$ and $k$. Eq.~\eqref{eq:bond_count} may be viewed as creating a histogram for each pair of elements. For feature vector \fv{2}, the histogram has just one bin and reports the number of bonds between those elements in the molecule. For feature vector \fv{2^{B}}, the number of bins in the histogram is the number of allowed bond orders between those elements. 

A simple approach to adding more information to the feature vectors is to add more bins to the histogram. The information encoded can also be expanded to include distances between non-bonded atoms. To accomplish this, we create a uniformly spaced grid,
\begin{align}
  d_i &= r^{start} + \frac{(r^{end}-r^{start}) i}{N_{grid}-1} \\  & \text{for}\ i = 0,1,\ldots,N_{grid}-1 \ \nonumber .
\end{align}
For each pair of elements, we then get a set of $N_{grid}$ features,
\begin{equation}\label{eq:hist}
v_{i} = \sum_{j<k} \mathbbm{1}(r_{jk} \in (d_i,d_{i+1}] ) \ .
\end{equation}
The feature vector of Eq.~\eqref{eq:hist} has two disadvantages. The first is that a small change in $r_{jk}$ can lead to a discontinuous change in the feature vector when $r_{jk}$ is near a grid point, $d_i$. In addition, for values of $N_{grid}$ that are large compared to the number of pairs of elements in the molecules of interest, the feature vector becomes sparse, making the learning prone to overfitting.

To overcome these disadvantages of the histogram-like features, we generalize Eq.~\eqref{eq:hist} by replacing $\mathbbm{1}(r_{jk} \in (d_i,d_{i+1}] )$ of Eq.~\eqref{eq:hist} with an encoding function, $f$,
\begin{equation}\label{eq:bond_weight}
v_{i} = \sum_{j<k} f(r_{jk}, d_{i}, \beta) .
\end{equation}
$f$ returns a value in the range $[0, 1]$ and may depend on a parameter, $\beta$, that is used to alter the smoothness of the function. We consider two general classes of encoding functions: probability density functions (PDF) and cumulative distribution functions (CDF). For both of these, we consider three types of functions: the normal distribution, the logistic distribution, and the spike distribution (Table \ref{tab:functions}).

\begin{table}[]
\centering
\begin{tabular}{lll}
    Name    & Name2 & $f(r_{jk}, d_{i}, \beta)$ in Eq.~\eqref{eq:bond_weight} \\
    \hline
    \fv{2^{NP}} & Normal PDF      & $\exp(-x^2) $\\
    \fv{2^{NC}} & Normal CDF      & $ 1 + \text{erf}(x) $ \\
    \fv{2^{LP}} & Logistic PDF    & $ \exp(-x) / (1 + \exp(-x))^2 $ \\
    \fv{2^{LC}} & Logistic CDF    & $1/(1 + \exp(-x))$ \\
    \fv{2^{SP}} & Spike PDF       & $\mathbbm{1}(|r_{jk}-d_{i}| < \epsilon / 2)$ \\
    \fv{2^{SC}} & Spike CDF       & $\mathbbm{1}(r_{jk} > d_{i})$ \\
\end{tabular}
\caption{Functions used for the encoded features of Eq.~\eqref{eq:bond_weight}. $r_{jk}$ is the distance between atoms, $d_i$ are the grid points, $\beta$ is a smoothing parameter, $x = \beta (r_{jk} - d_i)$, and $\epsilon$ is the distance between two grid points.}
\label{tab:functions}
\end{table}

For the PDF class of functions, the feature vector may be viewed as adding noise to each of the atom-atom distances, $r_{jk}$, with noise distributed according to the distribution of the encoding function. The CDF class of functions integrate the corresponding PDFs, leading to a feature vector that increases monotonically with distance and has a slope that indicates the presence of atom-atom separations near $d_i$.

\begin{figure*}
    \centering
    \includegraphics[width=7in]{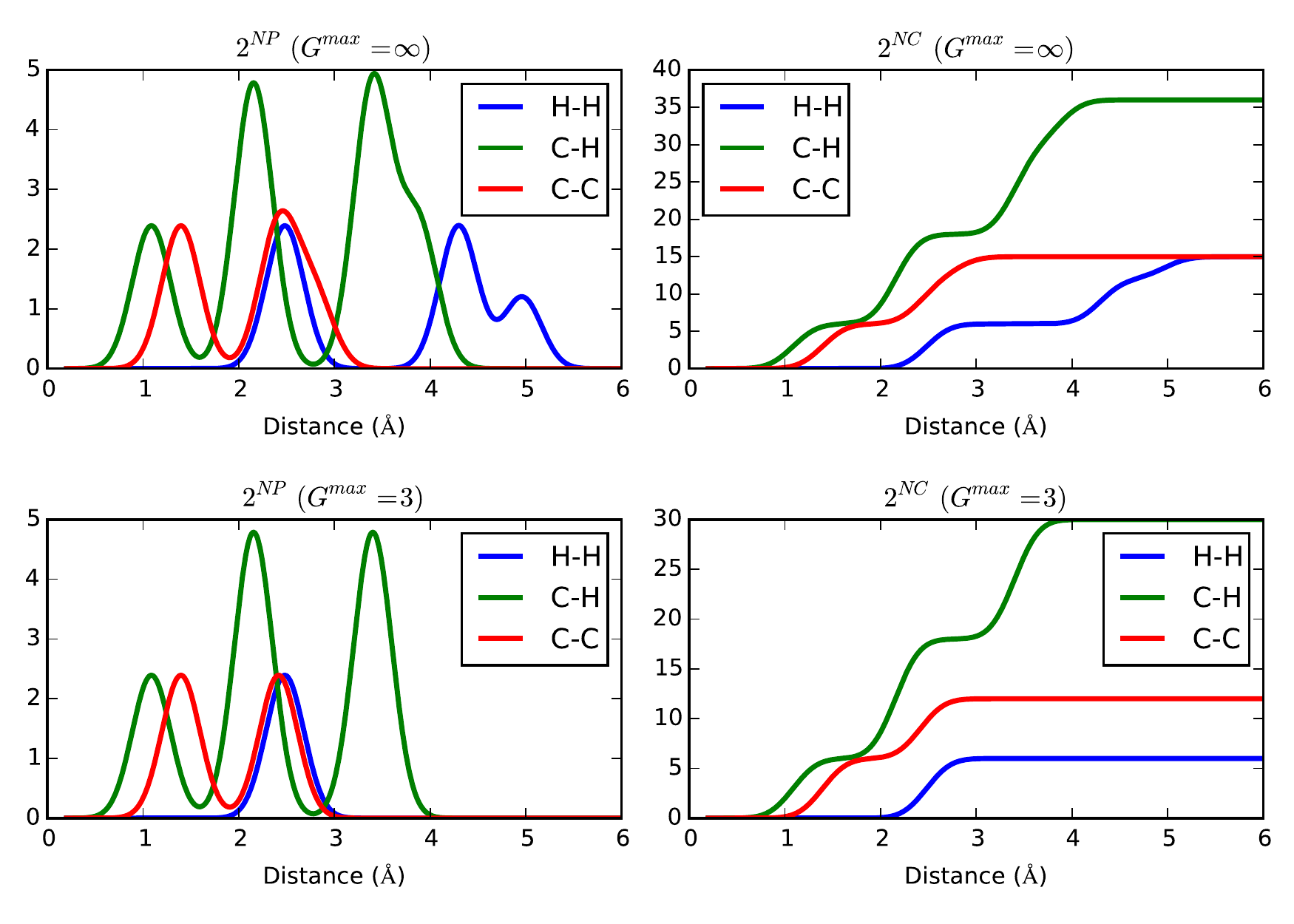}
    \caption{Encoded bond feature vectors for benzene. Left panels are \fv{2^{NP}} and right panels are \fv{2^{NC}}. Lower panels limit the geodesic distance between atoms, $G^{max}$ of Eq.~\eqref{eq:gmax}, to 3. Upper panels include all distances.}
    \label{fig:encoded}
\end{figure*}

Empirical results, discussed below, suggest that model performance may be improved by limiting the geodesic distance, $G$, between atoms included when computing the feature,
\begin{equation}\label{eq:gmax}
v_{i} = \sum_{j<k} f(r_{jk}, d_{i}, \beta) \mathbbm{1}(G_{jk} \le G^{max})
\end{equation}
where  $G_{jk}$ is an integer specifying the number of bonds along the shortest path between the atom $j$ and atom $k$. The last term limits the summation to pairs of atoms that are separated by up to $G^{max}$ bonds. Other works have taken a similar approach by setting a cutoff based on the Euclidean distance between atoms \cite{parrinello,nn_review,gastegger2015,gastegger2016,local_coulomb}. In the current studies, model performance was found to be somewhat better when using a cutoff based on geodesic distance. Examples of encoded features for benzene can be seen in Figure~\ref{fig:encoded}. 

\begin{figure*}
    \centering
    \includegraphics[width=7in]{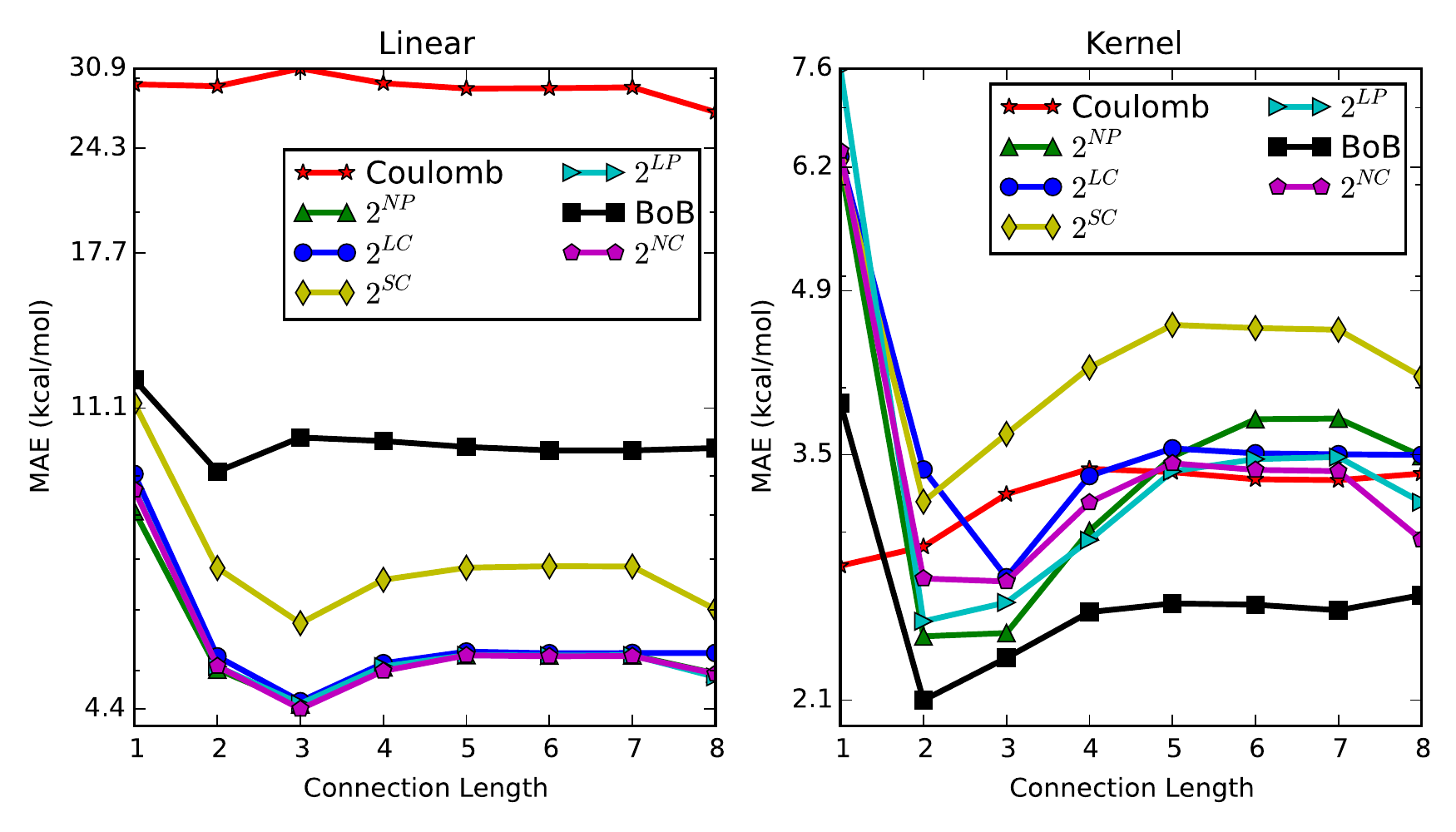}
    \caption{MAE of LRR (left) and KRR (right) models of QM7 atomization energies using each of the features listed in the legend, as a function of $G^{max}$ in Eq.~\eqref{eq:gmax}.}
    \label{fig:gmax}
\end{figure*}

The parameters specifying the encoded distance features were tuned by evaluating performance on the QM7 data set. The Supporting Information discusses the choice of $N_{grid}$ and $\beta$ of Eq.~\eqref{eq:bond_weight}.  For the results reported below, $\beta$ was $20\ \text{\AA}^{-1}$, and the grid consists of 100 points between $0.2\ \text{\AA}$ and $6\ \text{\AA}$. The choice of cutoff distance, $G^{max}$ of Eq.~\eqref{eq:gmax}, is based on the results in Figure~\ref{fig:gmax}, which show how model performance varies with $G^{max}$ for various encoded features. Results for the Coulomb matrix and BoB features are also shown, in which case $G^{max}$ is implemented by setting to zero those values of the Coulomb matrix that correspond to atoms separated by a geodesic distance greater than $G^{max}$. For nearly all features, the best performance is obtained with $G^{max}$ of 2 or 3, corresponding to distances over which angles and dihedrals would be assigned to the molecular structure. This is true of all features, when LRR is used, and nearly all features, when KRR is used. The exceptions are KRR with \fv{2^{SP}} (results not shown as the MAE is greater than 50 kcal/mol) and KRR with the Coulomb matrix. Note, however, that when the Coulomb matrix elements are reordered to form BoB features vectors, the optimal $G^{max}$ is in the 2-3 range. The observation that $G^{max}$ is in this range for a broad range of feature types and for both KRR and LRR models suggests that the atomization energy has a local character, whereby the energy of a given bond is primarily a function of its local environment. For the remainder of this work, $G^{max}$ is set to 3 for the encoded features. To allow comparison with past work, the BoB features do not limit $G^{max}$. 

Because the KRR models outperform LRR by more than 0.5 kcal/mol, we also discuss only results from KRR. (Full results from both LRR and KRR are tabulated in the Supporting Information.)

\section{Results}\label{sec:results}

Throughout this section, we summarize results by listing the features, with notation as in Section~\ref{sec:features}, followed in parentheses by the MAE of a KRR model developed as in Section~\ref{sec:ml}. For example, \fv{2^B} (7.69 kcal/mol) indicates that a KRR model using a feature vector \fv{2^B} leads to an MAE of 7.69 kcal/mol. The Supporting Information contains an extensive list of results from both LRR and KRR models based on different feature sets. Selected results are discussed here. 

We first consider models that use a single type of feature from Section~\ref{sec:features} to predict the atomization energies of the QM7 data set. For connectivity counts, the best performing single features are \fv{2^B} (7.69 kcal/mol) and \fv{3^{B}} (6.36 kcal/mol). The use of the 3D molecular structure via encoded distance features lowers the error substantially, with the best feature being \fv{2^{NP}} (2.45 kcal/mol). The other encoding functions perform nearly as well, except for \fv{2^{SP}} (146 kcal/mol) which is essentially a histogram and so gives a discontinuous and sparse feature vector. This suggests that smooth encoding functions lead to enhanced performance.

\begin{figure}[h]
    \centering
    \includegraphics[width=3.25in]{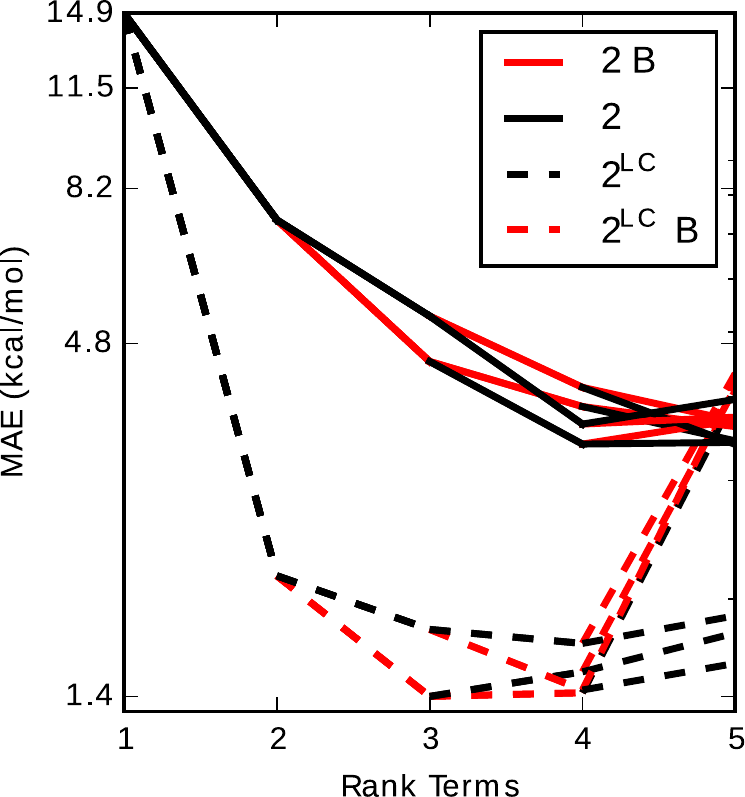}
    \caption{Effects of adding higher-rank features to concatenated feature vectors for models of QM7 atomization energies. Solid lines are features beginning with \fv{1 2^{B}}, requiring only atom connectivity. Dotted lines begin with \fv{1 2^{LC}} and thus encode distances from 3D geometries. For ranks 3 and higher, red lines show addition of features that distinguish bond order (\fv{3^{B}}, \fv{4^{B}}, or \fv{5^{B}}) versus simply bond existence (\fv{3}, \fv{4}, or \fv{5}).}
    \label{fig:adding_rank}
\end{figure}

Model performance can be enhanced by concatenating features of different ranks. The effects of adding features of increasing rank are shown in Figure~\ref{fig:adding_rank}, with selected results shown in Table~\ref{tab:energies}. The upper branch in Figure~\ref{fig:adding_rank} (solid lines) shows results when only connectivity counts are included and all feature vectors begin with \fv{1 2^{B}}. These have the advantage of requiring only a line drawing or SMILES string of the molecule. Addition of higher-rank connectivity counts improves performance up though rank 4, with addition of rank 5 leading to a small degradation. Figure~\ref{fig:adding_rank} also examines whether inclusion of bond order improves performance, e.g. whether \fv{3^{B}} (red line) or \fv{3} (black line) leads to a lower MAE. For feature vectors including only connectivity counts, inclusion of bond order enhances performance at rank 3, but degrades performance at higher ranks. The lowest MAE for concatenated connectivity counts is \fv{1 2^{B} 3^{B} 4} (3.40 kcal/mol).  

The lower branch (dashed lines) of Figure~\ref{fig:adding_rank} shows results from feature vectors that begin with \fv{1 2^{LC}}, and so include encoded distances from 3D molecular structures. For feature vectors that include encoded distances, inclusion of rank 3 leads to substantial improvements in model performance and inclusion of rank 5 degrades performance. Inclusion of rank 4 either slightly improves or slightly degrades performance, depending on both the type of rank-2 encoding employed and the property being predicted. In contrast to the connectivity only features, inclusion of bond order leads to better performance at both rank 3 and rank 4. The best performance on QM7 atomization energies is \simplef~(1.19 kcal/mol). Inclusion of rank 4, as \bestf~(1.25 kcal/mol), slightly degrades performance. The performance is significantly better than the  2.40~kcal/mol from BoB, which was the previous state-of-the-art for the QM7 data set\cite{bag_of_bonds}. 

\begin{table*}[]
\centering
\begin{tabular}{l|r|*{4}{r}}
        & & & \multicolumn{3}{c}{First $X$ thousand} \\
        & & & \multicolumn{3}{c}{molecules of QM9} \\
    feature & QM7 & CV & 20 & 50 & 133 \\
    \hline
    Null                    & 179.01 & 189.45 & 290.57 & 322.72 & 425.12 \\
    Coulomb matrix          &   3.37 & 3.83 & 43.78  & 66.56 & 106.09 \\
    BoB            &   2.40 & 2.43 & 9.68 & 22.21 & 29.99  \\ 
    \hline
    \fv{1}                     &  14.58 & 15.46 & 18.20 & 21.65 & 20.99 \\
    \fv{2^{B}}                 &   7.69 & 7.52 & 10.42 & 16.67 & 15.73 \\
    \fv{3^{B}}                 &   6.36 & 6.84 & 8.26 & 15.39 & 15.73 \\
    \fv{1 2^{B}}               &   6.88 & 6.37 & 8.75 & 11.11 & 11.17 \\
    \fv{1 2^{B} 3}             &   5.28 & 4.19 & 6.53  & 10.03  & 9.99 \\
    \fv{1 2^{B} 3^{B}}         &   4.51 & 4.25 & 6.34 & 8.36 & 8.59  \\
    \fv{1 2^{B} 3 4}           &   3.64 & 3.68 & 5.70  & 9.18 & 9.89  \\
    \fv{1 2^{B} 3 4^{B}}       &   4.13 & 3.81 & 6.10 & 9.90  & 10.51 \\
    \fv{1 2^{B} 3^{B} 4}       &   3.40 & 3.56 & 5.43  & 8.64 & 9.42  \\
    \fv{1 2^{B} 3^{B} 4^{B}}   &   3.87 & 3.69 & 5.92  & 9.53  & 10.25  \\ 
    \hline
    \fv{2^{LC}}                &   2.74 & 2.10 & 3.26 & 5.83 & 5.76 \\
    \fv{2^{NP}}                &   2.45 & 1.65 & 2.65 & 5.18 & 4.71 \\
    \fv{1 2^{LC}}              &   1.68 & 1.57 & 2.42  & 4.20 & 4.09 \\
    \fv{1 2^{LC} 3}            &   1.61 & 1.44 & 2.32  & 4.13  & 3.95  \\
    \fv{1 2^{LC} 3^{B}}        &   1.43 & 1.28 & 2.03  & 3.72 & 3.60  \\
    \fv{1 2^{LC} 3 4}          &   1.73 & 1.38 & 2.21 & 4.36 & 4.21 \\
    \fv{1 2^{LC} 3 4^{B}}      &   1.46 & 1.34 & 1.89 & 3.77  & 3.52  \\
    \fv{1 2^{LC} 3^{B} 4}      &   1.56 & 1.24 & 1.98 & 3.93  & 3.77  \\
    \fv{1 2^{LC} 3^{B} 4^{B}}  &   1.45 & 1.31 & 1.83  & 3.66  & 3.41  \\
    \fv{1 2^{NP}}              &   1.45 & 1.14 & 2.03 & 3.84 & 3.51 \\
    \fv{1 2^{NP} 3^{B}}        & \textbf{1.19} & \textbf{1.05} & 1.63 & \textbf{3.37} & \textbf{3.05} \\
    \fv{1 2^{NP} 3^{B} 4^{B}}  &   1.25  & 1.16 & \textbf{1.62} & 3.44  & 3.30  \\ 
    \hline
\end{tabular}
\caption{MAE of atomization energies for KRR models that use the listed feature vectors, for both the QM7 and QM9 data sets. The QM9 model is trained to molecules with 7 or fewer heavy atoms (3,993 molecules), yielding the cross validation error listed under CV. The remaining columns apply the model to subsets of QM9 with increasing molecular size. The lowest MAE for each column is shown in bold.}
\label{tab:energies}
\end{table*}

We next use the QM9 data set to explore the extent to which models trained on smaller molecules may be transferred to larger molecules. KRR models, using the features of Table~\ref{tab:energies}, were trained to the 3,993 molecules of the QM9 data set that have seven or fewer heavy atoms. Two-fold cross validation was used to select hyperparameters and the average MAE of the two folds, computed using the selected hyperparameters, is listed in the column labeled CV (for cross validation) in Table~\ref{tab:energies}. The remaining columns list the MAE as molecules from the QM9 data set are added to the small molecules used to train the model. Because the molecules are sorted by size, the average molecular size increases as molecules are added.  A useful quantitative measure of this increase is provided by the MAE of the null model, which reports on the spread of the atomization energies. The MAE of the null model for the entire QM9 data set is 2.4 times that of the smaller molecules of QM7. 

We first consider models that use only connectivity features. For simple models based on features that count elements (feature \fv{1}) or bond types (feature \fv{2^B}) the MAE for QM9 is 1.4 and 2.0 times that of QM7, respectively. This is less than the factor of 2.4 seen for the null model (Table~\ref{tab:energies}) and suggests that the average atomization energy per element, or average strength of a given bond type, does not change substantially with molecular size. Inclusion of higher-rank connectivity counts substantially improves model performance, with \fv{12^B3^B4} leading to the best performance on both QM7 and QM9. With \fv{12^B3^B4}, the MAE of QM9 is 2.8 times that of QM7, only slightly larger than the 2.4 increase in the null model. 

Not surprisingly, models that use the Coulomb matrix and BoB features do not transfer well to larger molecules. The MAEs of such models increase by over a factor of ten as one moves from small molecules to the entire QM9 data set. In addition, the MAEs on the full QM9 data set are larger than those from models based either on atom counts, \fv{1}, or bond counts, \fv{2^B}. 

Good transfer to larger molecules is, on the other hand, obtained for models based on encoded distance features. For models based on the \fv{2^{LC}} or \fv{2^{NP}} feature alone, the MAE for QM9 relative to QM7 increases by factors of 2.1 and 1.9, respectively, which is comparable to or less than the 2.4 increase of the null model. This advantage is retained upon inclusion of connectivity features of different rank. For both \simplef and \bestf, the MAE of QM9 is 2.6 times that of QM7, which is only slightly larger than the 2.4 increase in the null model. 

\begin{table*}[]
\centering
\begin{threeparttable}
\begin{tabular}{lrrrrrrr}
    \hline
    \hline
    & \multicolumn{6}{c}{Mean Absolute Errors} \\
    Property        & Null & \simplef & \bestf        &  CM\cite{coulomb_multi_prop} & BoB\tnotex{tnote:qm7b-10}  & SOAP\cite{resoap} & QM\\
    \hline
    E (PBE0) (eV)                       & 7.69 & \textbf{0.04} &         0.05  &          0.16 &       0.08 & \textbf{0.04} & 0.09-0.23\tnotex{tnote:qm7b-1}    \\
    $\alpha$ (PBE0) ($\text{\AA}^3$)    & 1.04 &         0.06  &         0.07  &          0.11 &       0.08 & \textbf{0.05} & 0.04-0.27\tnotex{tnote:qm7b-2}    \\
    $\alpha$ (SCS) ($\text{\AA}^3$)     & 1.15 &         0.05  &         0.06  &          0.08 &       0.04 & \textbf{0.02} & 0.04-0.27\tnotex{tnote:qm7b-2}    \\
    HOMO (GW) (eV)                      & 0.54 &         0.13  &         0.14  &          0.16 &       0.14 & \textbf{0.12} & --                                \\
    HOMO (PBE0) (eV)                    & 0.49 &         0.12  &         0.12  &          0.15 &       0.13 & \textbf{0.11} & 2.08\tnotex{tnote:qm7b-6}                     \\
    HOMO (ZINDO) (eV)                   & 0.78 & \textbf{0.13} & \textbf{0.13} &          0.15 &       0.14 & \textbf{0.13} & 0.79\tnotex{tnote:qm7b-7}                     \\
    LUMO (GW) (eV)                      & 0.31 &         0.13  &         0.13  &          0.13 &       0.15 & \textbf{0.12} & --                       \\
    LUMO (PBE0) (eV)                    & 0.53 &         0.09  &         0.09  &          0.12 &       0.11 & \textbf{0.08} & 1.30\tnotex{tnote:qm7b-7}                     \\
    LUMO (ZINDO) (eV)                   & 1.08 & \textbf{0.10} & \textbf{0.10} &          0.11 &       0.14 & \textbf{0.10} & 0.93\tnotex{tnote:qm7b-7}                     \\
    IP (ZINDO) (eV)                     & 0.78 & \textbf{0.16} &         0.18  &          0.17 &       0.19 &         0.19  & 0.20, 0.15\tnotex{tnote:qm7b-4}                \\
    EA (ZINDO) (eV)                     & 1.17 & \textbf{0.09} &         0.11  &          0.11 &       0.15 &         0.13  & 0.16\tnotex{tnote:qm7b-8}, 0.11\tnotex{tnote:qm7b-4}               \\
    $E^*_{1st}$ (ZINDO) (eV)            & 1.54 &        0.23   &         0.28  &  \textbf{0.13}&       0.20 &         0.18  & 0.18\tnotex{tnote:qm7b-8}, 0.21\tnotex{tnote:qm7b-9}               \\
    $E^*_{max}$ (ZINDO) (eV)            & 2.62 &        1.23   &         1.27  &  \textbf{1.06}&       1.30 &         1.56  & --                       \\
    $I_{max}$ (ZINDO) (Arb.)            & 0.15 & \textbf{0.06} &         0.07  &          0.07 &       0.08 &         0.08  & --                       \\
    \hline
    \hline
\end{tabular}
    \begin{tablenotes}

      \item\label{tnote:qm7b-1}
      0.15 eV from PBE0, MAE of formation enthalpy for the G3/99 set\cite{staroverov2003comparative,curtiss2005assessment}.\\
      0.23 eV from PBE0, MAE of atomization energy for six small molecules\cite{zhao2004tests,lynch2003small}.\\
      0.09-0.22 eV from B3LYP, MAE of atomization energy from various studies\cite{koch2015chemist}.

      \item\label{tnote:qm7b-2}
      0.05-0.27 $\text{\AA}^3$ from B3LYP, MAE from various studies\cite{koch2015chemist}.\\
      0.04-0.14 $\text{\AA}^3$ from MP2, MAE from various studies\cite{koch2015chemist}.
      \item\label{tnote:qm7b-4}B3LYP, MAE from various studies\cite{koch2015chemist}.
      \item\label{tnote:qm7b-6}MAE from GW values\cite{coulomb_multi_prop}.
      \item\label{tnote:qm7b-7}ZINDO, MAE for a set of 17 retinal analogues\cite{lopez2006computation}.
      \item\label{tnote:qm7b-8}PBE0, MAE for the G3/99 set\cite{staroverov2003comparative,curtiss2005assessment}.
      \item\label{tnote:qm7b-9}TD-DFT(PBE0), MAE for a set of 17 retinal analogues\cite{lopez2006computation}.
      \item\label{tnote:qm7b-10}Calculations done in this work for comparison.
    \end{tablenotes}
\end{threeparttable}
\caption{Comparison of models for the properties in the QM7b data set. CM and SOAP lists MAEs reported for the Coulomb Matrix\cite{coulomb_multi_prop} and REMatch-SOAP\cite{resoap}. BoB is from models developed here using BoB features as in Section~\ref{sec:ml}. \simplef and \bestf~list MAEs from models developed using feature vectors of the type developed in this work. QM lists the estimated reliability of the quantum chemical calculations used to create the QM7b data set.\cite{coulomb_multi_prop}}
\label{tab:qm7b}
\end{table*}

The degree to which various feature vectors are useful for creating models of other target properties is explored using the QM7b data set. Table~\ref{tab:qm7b} lists the MAE from BoB models developed here, along with results reported from use of the Coulomb matrix \cite{coulomb_multi_prop} and REMatch-SOAP kernel\cite{resoap}. These are compared to models developed using \simplef, which gives the best performance on QM7 atomization energies (Table~\ref{tab:energies}), and \bestf, which extends this to rank 4 and gives comparable results on QM7b. For the remaining studies reported below, which include larger molecules and larger data sets, we use the longer \bestf\ feature. The results show that the performance with \simplef\ and \bestf\ is comparable to BoB and REMatch-SOAP, all of which lead to predictions whose MAEs are at least as low as the typical deviation of the quantum methods from experiment used to generate the QM7b data set (QM in Table~\ref{tab:qm7b}). 

\begin{figure}[h]
    \centering
    \includegraphics[width=3.25in]{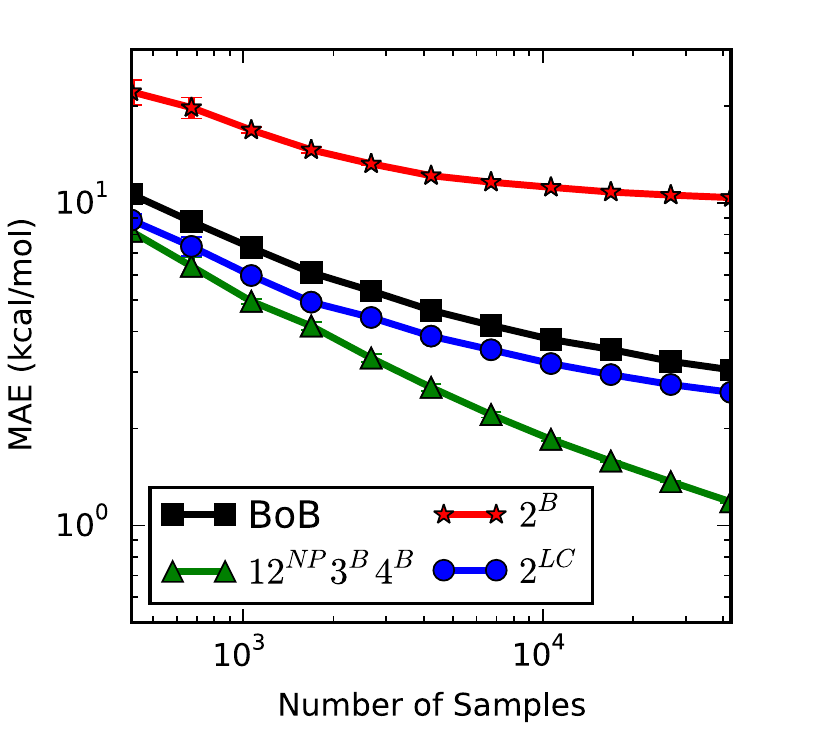}
    \caption{MAE of atomization energy predictions, as a function of the number of examples included from the QM9 data set. The error bars, which are often smaller than the line width, show the standard deviation from five experiments, each of which sampled molecules at random from the full data set. Similar learning rates are obtained for all feature vectors.  (Hyperparameters are listed in the Supporting Information.)
    }
    \label{fig:qm9_sample}
\end{figure}

\begin{figure*}
    \centering
    \includegraphics[width=7in]{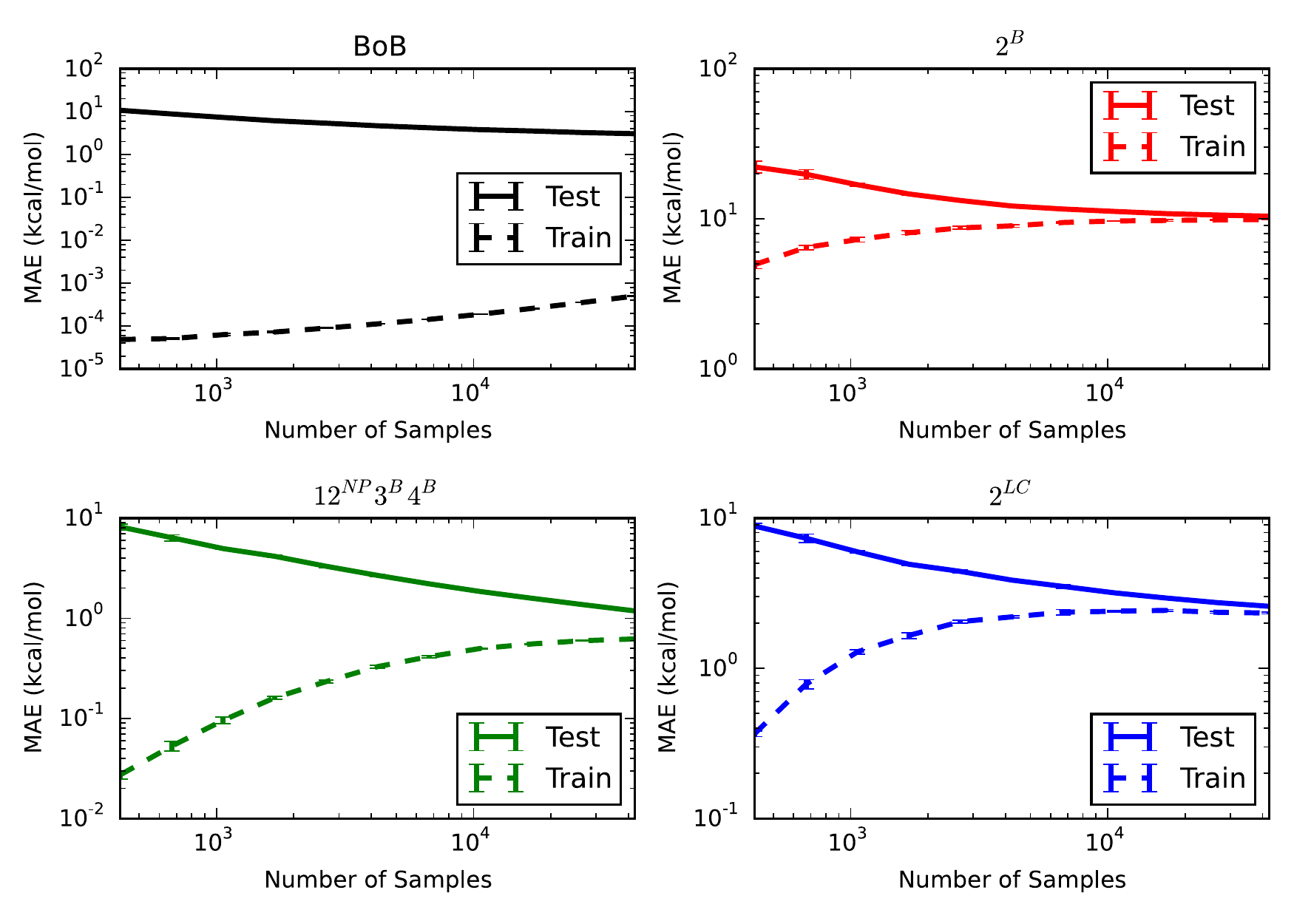}
    \caption{Comparison of test and train errors as a function of number of examples included from the QM9 data set. The error bars, which are often smaller than the line width, show the standard deviation from five experiments, each of which sampled molecules at random from the full data set.}
    \label{fig:rate}
\end{figure*}

The learning rate of models based on various feature vectors are shown in Figure~\ref{fig:qm9_sample}. Similar learning rates are obtained across feature vectors, with the decrease in MAE for models based on simple bond counting, \fv{2^B}, being similar to those of BoB, \fv{2^{LC}}, and \bestf. Figure~\ref{fig:rate} provides additional insight by showing the MAE for both the training and test data. The point at which the train and test error converge indicates the limit beyond which inclusion of additional data may no longer be expected to improve model performance. The model based on the concatenated \bestf~feature approaches this convergence significantly more slowly than the single \fv{2^{B}} or \fv{2^{LC}} features, suggesting that the increase in performance of \bestf~comes at the cost of a somewhat slower learning rate. For BoB, the training error is nearly zero, suggesting that inclusion of additional data has the potential to further reduce error.

\begin{table*}[]
\centering
\begin{tabular}{l|*{7}{r}}
        &  & \multicolumn{6}{c}{Sampled Molecules} \\
    Property &  Null &  500 &   1000 &  2000 &  4000 &  8000 &  16000 \\
    \hline
$\mu$ (Debye)       &  1.19    &   0.92(1)  &   0.86(1)  &   0.83(1)  &   0.76(1)  &   0.68(0)  &   0.63(0)   \\
$\alpha$ (Bohr$^3$) &  6.30    &   0.98(4)  &   0.77(2)  &   0.66(1)  &   0.60(4)  &   0.49(2)  &   0.41(1)   \\
HOMO (eV)           &  0.44    &   0.25(0)  &   0.22(1)  &   0.19(0)  &   0.17(0)  &   0.15(0)  &   0.12(0)   \\
LUMO (eV)           &  1.05    &   0.39(1)  &   0.32(2)  &   0.26(0)  &   0.22(0)  &   0.18(0)  &   0.15(0)   \\
Gap (eV)            &  1.08    &   0.44(0)  &   0.39(2)  &   0.32(0)  &   0.28(0)  &   0.24(0)  &   0.19(0)   \\
$\langle R^2 \rangle$ (Bohr$^2$)    &  202.52  &  86(3)  &  71(2)  &  59.41(91)  &  50.40(54)  &  44.94(35)  &  36.31(19)   \\
zpve (kcal/mol)     &  16.63   &   0.29(1)  &   0.21(1)  &   0.17(0)  &   0.14(0)  &   0.11(0)  &   0.09(0)   \\
U$_0$ (kcal/mol)    &  188.47  &   7.28(33)  &   4.67(10)  &   3.64(13)  &   2.75(5)  &   2.08(5)  &   1.58(2)   \\
U (kcal/mol)        &  190.18  &   7.31(33)  &   4.69(10)  &   3.63(11)  &   2.77(5)  &   2.09(5)  &   1.59(2)   \\
H (kcal/mol)        &  191.55  &   7.32(33)  &   4.69(10)  &   3.63(11)  &   2.77(5)  &   2.09(5)  &   1.59(2)   \\
G (kcal/mol)        &  173.30  &   7.08(31)  &   4.58(7)  &   3.57(11)  &   2.73(4)  &   2.06(5)  &   1.56(2)   \\
C$_v$ (cal/(mol K))    &  4.95    &   0.32(0)  &   0.24(0)  &   0.20(0)  &   0.16(0)  &   0.12(0)  &   0.10(0)   \\
\end{tabular}
\caption{MAEs for KRR models that use the \bestf~feature vector to target properties in the QM9 data set. The null column indicates the spread of the data. The columns are for training on X molecules and testing on the remainder. The uncertainties are the standard deviations from four experiments that sampled the X training molecules at random from the entire data set.}
\label{tab:qm9props}
\end{table*}

The performance and learning rate of models, based on the \bestf~feature vector, that target a range of properties in the QM9 data set are shown in Table~\ref{tab:qm9props}. Performance and learning rate on all thermodynamic energies are roughly equivalent, giving MAEs of 1.6 kcal/mol. The heat capacity is also well predicted, with an MAE of 0.10 cal/(mol K) that is 2\% that of the null model. Models of the HOMO and LUMO energies, and the HOMO-LUMO gap, also lead to reasonable performance, with MAEs below 0.2~eV. 

The performance on dipole moment, $\mu$, is relatively poor, with the \bestf~model reducing the MAE of the null model by only about one half. This may be due to $\mu$ having a strong dependence on the overall geometry of the molecule. Significantly better performance is obtained for the polarizability, $\alpha$, where MAE of the \bestf~model is 6.5\% that of the null model. Although polarizability is a global property, it is primarily dependent on the total volume of the molecule and thus is likely well modeled as a sum of fragment properties. The dipole moment, on the other hand, depends on the detailed global arrangement of the atoms.

\section{Discussion}

This paper introduces feature vectors for machine learning of molecular properties whose length depends on the diversity of molecules in the data set, e.g. the number of elements and bond types, but is independent of the size of the molecules. These features lead to models whose performance on atomization energies, and other properties tabulated in well studied data sets, is comparable to or better than previous models based on the Coulomb matrix or BoB features. In contrast to these previous models, the features introduced here allow models trained on small molecules to be applied successfully to larger molecules. 

Although models based solely on connectivity counts would be ideal, because they require only the information present in a SMILES string, such models do not perform well. This reflects the fact that chemistry is more complex than simple Lewis structures---even for organic molecules. When information regarding the 3D geometry of the molecule is included via encoded distance features, the model performance improves substantially and this performance enhancement is retained upon transfer from smaller to larger molecules. 

The current work focuses on KRR models, for which the features are used to compute distances between pairs of molecules. These distances are then passed through a kernel for use in the regression. The above results suggest that the features developed here are useful for computing distances between molecules, when coupled to the standard kernels employed in machine learning. The development of kernels that better describe molecular similarity is left to future work. The degree to which the features introduced here lead to improved performance in deep learning and other ML models also remains to be explored; although, we note that the models with the best current performance on the data sets studied here use KRR. 

\section{Acknowledgements}

The authors thank Haichen Li for helpful discussions. The authors would also like to thank Raghunathan Ramakrishnan for helpful discussions on details regarding the BoB model and the data sets studied here. This work was supported by the National Science Foundation under grant CHE-1027985.

\section{Supporting Information Available}

Supporting information includes the following: distances between atoms used to define the bond types of Section~\ref{sec:conn}; details regarding the implementation of the encoded distance features of Section~\ref{sec:encoded}; and a listing of the MAE and hyperparameters from LRR and KRR models of the QM7 data set, obtained using a wide variety of different feature vectors.

\bibliographystyle{achemso}
\bibliography{main}

\end{document}